# Few-cycle pulse-driven excitation response of resonant medium with nonlinear field coupling

A V Pakhomov[1,2], R M Arkhipov[3,4], I V Babushkin[5,6] and M V Arkhipov[4]

[1] Department of Physics, Samara University, Moskovskoye Shosse 34, Samara 443086, Russia
[2] Department of Theoretical Physics, Lebedev Physical Institute, Novo-Sadovaya Str. 221, Samara 443011, Russia
[3] ITMO University, Kronverkskiy prospekt 49, St. Petersburg 197101, Russia
[4] Faculty of Physics, St. Petersburg State University, Ulyanovskaya 1, St. Petersburg 198504, Russia
[5] Institute of Quantum Optics, Leibniz University Hannover, Welfengarten 1, 30167, Hannover, Germany
[6] Max Born Institute, Max-Born-Strasse 2a, 10117, Berlin, Germany

E-mail: antpakhom@gmail.com

**Abstract.** We demonstrate that the resonant medium with essentially nonlinear field coupling can exhibit specific response when excited by the few-cycle optical pulse. We provide an effective theoretical background that allows relating the type of the field coupling function with the medium oscillators output on the few-cycle pulse-driven excitation. Possible applications of such the optical response are discussed for the controllable generation and shaping of the unipolar videopulses.

## 1. Introduction

Recent progress in generation of femtosecond and subfemtosecond optical pulses has brought up multiple questions concerning their interaction with different mediums [1-4]. Single and even half optical cycle duration of such pulses is comparable with the intramolecular and intraatomic processes timescales, thus opening up new fields of their use for the time-resolved control of ultrafast processes in matter [5-9]. Therefore, investigation of new laser-matter interaction phenomena caused by the few-cycle pulses turns to be a challenging area of the scientific research.

Study of the medium optical response on such the extremely-short pulses helps to reveal the mechanisms of their interactions with the medium, thus enabling to unravel the structure of matter. A lot of new physical effects related with the medium optical response are discovered for the intense few-cycle pulses when the pulse field strength becomes comparable with the intraatomic fields. Such the strong fields are able to transform the medium structure itself leading to novel phenomena of light-matter interaction like the multiphoton ionization, high-harmonic generation and femtosecond filamentation [10-14]. However, some interesting effects originating from the ultrashort pulse duration can also arise for even weak electromagnetic field, enabling the classical medium description to be valid. In prior paper [15], few-cycle pulse-driven excitation of the resonant medium composed of classical harmonic oscillators was examined and unusual properties of the transient medium radiation were found. It was demonstrated that analysis of the spectral and temporal characteristics of the medium response allows to obtain the information about the resonant medium properties. The control of this radiation parameters by the train of ultra-short pulses was examined in [16]. In recent paper [17], the optical response of the Raman-active particles excited by the few-cycle optical pulses was studied. A novel possibility of unipolar videopulse generation was predicted under excitation by a train of few-cycle pulses with proper time delay. Due to unique property of unidirectionality, the usage of such unipolar pulses opens new opportunities to steer the charges dynamics in matter. Hence, finding the efficient methods of their controllable generation seems to be challenging.

In this Letter, we provide a more detailed theoretical analysis of the optical response of resonant medium having the general form of nonlinear field coupling and being excited by the few-cycle optical pulse. We develop the proper theoretical underpinning and find out the specifics of the nonlinear oscillator

Few-cycle pulse-driven excitation response of resonant medium with nonlinear field coupling response on the few-cycle pulse-driven excitation. In contrast to the linear oscillator case, the nonlinear field coupling was shown to provide the novel effective way for the unipolar videopulse generation in the wide frequency range by all-optical means using a pair of successive few-cycle light pulses. We also propose a new method to control the videopulse waveform by interference of the multiple oscillators emission with excitation velocity varying in a definite manner along the oscillators array, contrary to the previously studied in [17] excitation with constant velocity.

**2. The effect of the field coupling function on the medium optical response**

Response of the oscillator to the external field can be generally described by the induced medium polarization $P(t)$. We consider the medium interacting with the low-intensity field, thus taking a model of the classical optical oscillator, but being driven by the exciting pulse field in the generalized form:

$$\ddot{P} + \gamma \dot{P} + \omega_0^2 P = g[E(t)] \cdot E(t), \tag{1}$$

where $\omega_0$ is the oscillator resonant frequency, $\gamma$ is the media damping rate and the function $g[E(t)]$ describes the media coupling strength to the field. We consider the exciting field linearly polarized, thus reducing the problem to the scalar one.

Let us assume the medium excited by the few-cycle pulse having the Gaussian pulse envelope:

$$E(t) = E_0 e^{\frac{-t^2}{\tau_p^2}} \sin(\Omega t). \tag{2}$$

The central frequency $\Omega$ is suggested to be much greater the medium resonant frequency $\omega_0$, so that the population dynamics of the excited levels can be neglected. We also suppose the pulse duration to be small compared to the period of resonant oscillations: $\omega_0 \tau_p \ll 1$.

Neglecting the oscillations decay during the exciting pulse duration and setting $u(t) = \dot{P}(t) + i\omega_0 P(t)$, we get the equation (1) in the form:

$$\dot{u} - i\omega_0 u = g[E(t)] \cdot E(t). \tag{3}$$

The solution of the equation (3), describing the polarization dynamics after the ultra-short pulse passing through, is:

$$u(t) = e^{i\omega_0 t} \left\{ u_0 + \int_{-\infty}^{+\infty} g[E(t')] \cdot E(t') e^{-i\omega_0 t'} dt' \right\}, \tag{4}$$

The integral in the right hand of equation (4) is taken over the whole pulse duration, what is schematically outlined in the infinite integration limits.

Let us first consider the conventional case of the oscillator being linear, implying that $g[E(t)] = g_0 = const$. This is the classical model commonly used to describe the bounded oscillating electron in the molecule. Given that $\omega_0 \tau_p \ll 1$ we can take in the expression under integral sign $e^{-i\omega_0 t'} \approx 1$, giving:

$$u(t) = e^{i\omega_0 t} \left\{ u_0 + g_0 \int_{-\infty}^{+\infty} E(t') dt' \right\}. \tag{5}$$

where $u_0$ is the integration constant describing the oscillator motion before the exciting pulse arrival.

Using the pulse equation (2), the integral in equation (5), that is proportional to the pulse area, becomes zero. Hence, we need here to keep the next term in the exponent expansion series with respect to the small parameter $\omega_0 t'$, taking $e^{-i\omega_0 t'} \approx 1 - i\omega_0 t'$. Upon that from the equation (4) we obtain:

$$u(t) = e^{i\omega_0 t} \left\{ u_0 - i\omega_0 g_0 \int_{-\infty}^{+\infty} E(t') t' dt' \right\}. \tag{6}$$

Few-cycle pulse-driven excitation response of resonant medium with nonlinear field coupling

Setting $\Pi = \omega_0 g_0 \int_{-\infty}^{+\infty} E(t')t' \mathrm{d}t'$ and separating real and imaginary parts in equation (6), we finally obtain the oscillator motion for $t > 0$ in the form:

$$P(t) = P_0 \sin(\omega_0 t + \phi_0) - \frac{\Pi}{\omega_0}\cos(\omega_0 t);$$
$$\dot{P}(t) = \omega_0 P_0 \cos(\omega_0 t + \phi_0) + \Pi \sin(\omega_0 t) \tag{7}$$

Let the oscillator be at a standstill for $t \leq 0$ before the exciting pulse arrival: $P_0 = 0$. Since $\omega_0 \tau_p \ll 1$ we can see from equation (7) that exciting few-cycle pulse induces an instant jump of the media polarization without changing its time derivative. This means that linear oscillator being excited by the optical pulse much shorter then its own oscillation period gets certain dipole moment without getting the dipole moment rate. It is important to note, that keeping higher order terms in exponent expansion series would not change this principal result.

We turn now to the case of medium with essentially nonlinear field coupling. We take the coupling strength to the field given as $g[E(t)] = \chi E(t)$. Such the field coupling is expected to happen for the nonlinearly coupled optical oscillators with strongly different parameters, like the effective mass and resonant frequency. Under such condition only one oscillator will be excited by the high-frequency external field, inducing in its turn the other oscillator motion through the nonlinear bonding. This can be potentially realized with the nonlinearly coupled localized plasmonic resonances in the metallic nanostructures [18-19], hybrid aggregates of organic supramolecular assemblies and inorganic nanocrystals [20-21], coupled semiconductor microcavities [22], double quantum dot heterostructures [23-24] or other hybrid optical materials. Thanks to their associated mature technologies the nonlinear optical properties of such complex structures can be engineered in a tunable way in accordance with the specified demand. The natural examples of similar field coupling include, without being limited to, the Raman-active medium since it is usually modeled with nonlinearly bonded electron and nucleus optical oscillators [25-26]. We shall consider in the following the nonlinear oscillator response in general terms without specifying its physical origin. Thereupon the equation (5) turns into:

$$u(t) = e^{i\omega_0 t}\left\{u_0 + \chi \int_{-\infty}^{+\infty} E^2(t')\mathrm{d}t'\right\}. \tag{8}$$

The integral in the right hand side $\widetilde{\Pi} = \chi \int_{-\infty}^{+\infty} E^2(t')\mathrm{d}t'$ is intrinsically nonzero since it is proportional to the pulse energy. With that equation (8) yields for $t > 0$:

$$P(t) = P_0 \sin(\omega_0 t + \phi_0) + \frac{\widetilde{\Pi}}{\omega_0}\sin(\omega_0 t);$$
$$\dot{P}(t) = \omega_0 P_0 \cos(\omega_0 t + \phi_0) + \widetilde{\Pi}\cos(\omega_0 t) \tag{9}$$

Let the oscillator be once again at a standstill before the exciting pulse arrival: $P_0 = 0$. In contrast to the linear case exciting ultra-short pulse doesn't change now the media polarization but induces an instant jump of its time derivative. That is the optical oscillator with nonlinear field coupling, being excited by the optical pulse much shorter then its own oscillation period, gets certain dipole moment rate without getting the dipole moment itself. It should be noted again, that keeping higher order terms in exponent expansion series would not change this type of response in no way. When dealing with the ultrafast processes, taking place on the timescales comparable with the oscillation period, the type of the medium response can be of principal importance. In the next section we demonstrate that the specific optical response in the nonlinear coupling case allows to enable the production of unipolar videopulses.

**3. Controlling the medium radiation by the train of few-cycle pulses**

# Few-cycle pulse-driven excitation response of resonant medium with nonlinear field coupling

Let us now consider how the second identical ultra-short pulse influences the oscillator motion. The oscillator response is again described by equations (7) and (9), but the first term now stands for the oscillations caused by the first exciting pulse. Provided $\gamma \ll \omega_0$, the far-field emitted by the oscillator at an arbitrary point $\bar{r}'$ is represented as:

$$\overline{E}(\bar{r}',t) \sim \overline{\ddot{P}}(\bar{r}, t - |\bar{r}-\bar{r}'|/c) \sim \omega_0^2 \overline{P}(\bar{r}, t - |\bar{r}-\bar{r}'|/c). \qquad (10)$$

Hence, except for the constant factor emitted field is again described by the equations (7), (9). Given that we can show that in medium with essentially nonlinear field coupling the unipolar videopulse generation becomes possible, that is the pulses with the nonzero time-averaged electric field $\int_{-\infty}^{+\infty} E(t)\mathrm{d}t \neq 0$, in contrast to common bipolar optical pulses with $\int_{-\infty}^{+\infty} E(t)\mathrm{d}t = 0$. This distinguishing feature of unipolar videopulses makes them promising to control and monitor ultrafast processes in real time, like the motion of charges in solid-state electronics or extremely nonlinear optics. The half-cycle videopulses have been previously obtained experimentally in terahertz range from a laser-driven plasma in a solid target [27]. Transformation of the ultra-short bipolar pulse into the unipolar one upon propagating in nonlinear media was also predicted theoretically by several authors [28-31].

As it directly follows from equation (10), to produce the unipolar videopulse one needs to stop the oscillator emission in the certain moment in such a way, that the in both final and initial moments of its oscillations oscillator has the same sign of deviation from the equilibrium. Findings of the previous section reveal that in linear case it turns impossible to implement: according to equation (7) to stop the oscillator we have to launch the second pulse at the moment of the maximum polarization amplitude in the opposite direction, what is possible just after changing the sign of the deviation.

The situation appears to be reverse for the oscillator with the nonlinear field coupling. According to equation (9), oscillator may be stopped after half-period (or odd number of half-periods) when its polarization value turns to zero while its rate reaches maximum value. For this purpose two ultra-short pulses separated by the time interval $T_\mathrm{p}$ equal to half-period $T_0/2$ (or odd number of half-periods) of natural oscillations should be launched in the medium. Since we did not impose any restrictions on the medium resonant frequency $T_0$, proposed method can be principally applied to the half-cycle pulse generation in the wide frequency range. It is worthwhile to say that this method is similar to the well known method of optical control over elementary molecular motion with sequences of femtosecond pulses, when one pulse initiates the required motion and second pulse can stop it [32].

Since a single oscillator generates just a half-sine videopulse (or odd number of half-sines), it doesn't allow to get the pulse with required spatio-temporal profile, what is of crucial importance for the optical coherent control applications. For this purpose, it seems appropriate to consider the summation of the videopulses generated by many single oscillators and brought together in the certain manner.

To illustrate the advantages of the proposed idea, we restrict ourselves in this Letter to the simplest case of one-dimensional oscillators string excited by two few-cycle optical pulses with the $T_0/2$ delay and cylindrical or spherical incident wavefronts, for instance from the point-light source (see figure 1). Hereafter we assume the exciting pulses linearly polarized with the polarization direction orthogonal to the plane of figure 1. Under such the assumption we get the problem of the single oscillator's emission interference to be scalar.

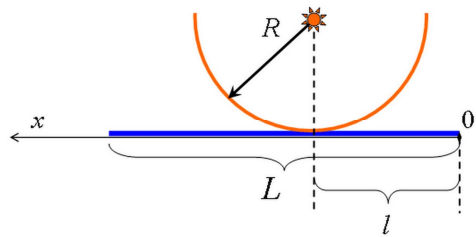

**Figure 1.** A one-dimensional string composed of optical oscillators with nonlinear field coupling is excited by two successive ultra-short light pulses with cylindrical or spherical

Few-cycle pulse-driven excitation response of resonant medium with nonlinear field coupling wavefronts. Radiation from the medium is observed in a far distant point on *x* axis with coordinate *r*.

Given equation (10) and assuming $\gamma \ll \omega_0$, mathematical expression for the generated pulse shape is written as:

$$E(t) = E_0 \sum_{k=0}^{1} \int_0^L \sin[\omega_0 f_x]\Theta[f_x]dx,\qquad(11)$$

where $f_x = t - \frac{r-x}{c} - \frac{\sqrt{R^2+|x-l|^2}-R}{c} - (k-1)T_p$, $\Theta$ is Heaviside step-function, $E_0$ is the scaling constant, $T_p = T_0/2$ is the pulse repetition period and the observation point is located on the string axis at the distance $r \gg L$.

The results of the numerical calculation of the integral (11) for different values of the parameters $b = \frac{\omega_0 L}{c}$, $R/L$, $l$ are plotted in figures 2(a) and (b). The profile of the generated videopulse turns to be monotonically decreasing from its highest amplitude at the leading edge to the lowest at the trailing edge. In such the one-dimensional configuration this shape results from the fact that the intersection point of exciting wavefront moves along the string with the superluminal velocity varying along the string due to wavefront curvature. This is in contrast to the excitation with constant velocity studied in [17] that permits generation of only flat-top pulses. The final shape asymmetry is determined by the character of this excitation velocity varying along the string leading to the condensation of single half-cycle pulses along the propagation towards the leading edge. Shifting the exciting wavefront center over the oscillators array allows just to slightly tune the pulse duration and its amplitude (see figure 2(b)). It is seen that selecting the oscillators geometrical composition and exciting pulses waveform allows in principle to obtain the unipolar pulses with widely tunable shape.

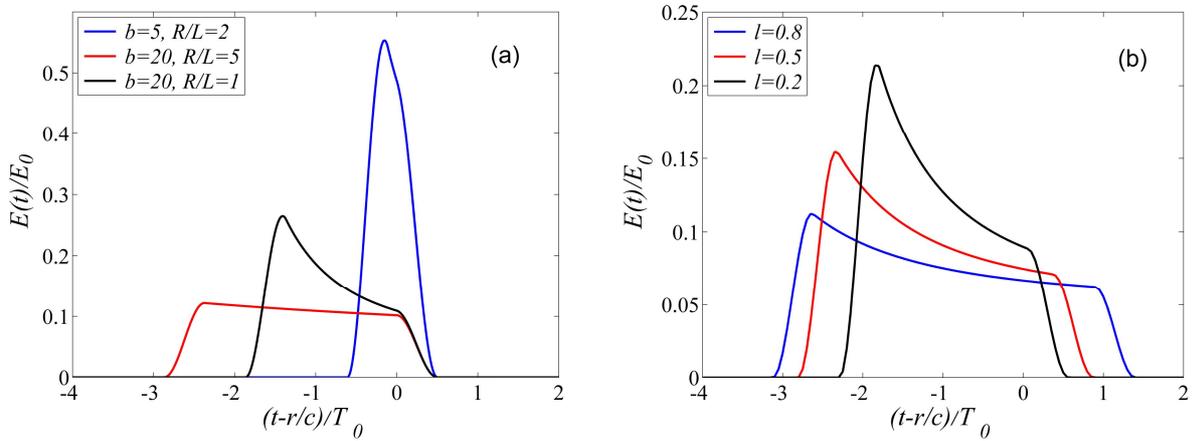

**Figure 2.** Results of the numerical solution of the integral (11) for different values of parameters *b, R/L, l*: (a) *l=0*; (b) *b=20, R/L=1*.

**4. Conclusion**
We studied theoretically the few-cycle pulse-driven exciting response of optical oscillator with the essentially nonlinear field coupling. It was shown to exhibit specific features that can not be obtained for the linear oscillator. Namely, while linear oscillator can be given just the certain dipole moment without getting the dipole moment rate, the situation turns to be quite opposite for the nonlinear oscillator with the coupling function proportional to the driving field. This specific exciting response was found to enable the unipolar videopulse generation, when the oscillator is influenced by the pair of pulses with the proper time delay, so that the first pulse initiates the videopulse emission and the second one stops it at the certain moment. This approach can be potentially extended to the half-cycle pulse generation in the wide frequency range, primarily when the optical period is well above the duration of femtosecond, like in the terahertz and mid-infrared range, which are lacking of the ultra-short pulse sources.

# Few-cycle pulse-driven excitation response of resonant medium with nonlinear field coupling

To fully exploit the potential of proposed approach for the unipolar videopulse generation and especially for the pulse shaping, we considered the interference of the videopulses emitted by a linear string of oscillators when irradiated by the exciting few-cycle pulses with curved wavefront. Depending on the geometry of oscillators composition and the wavefront curvature we can tune the resulting pulse shape within certain extent according to the desirable spatio-temporal parameters of the emitted pulse. It is important to stress that the development of techniques to control the ultra-short pulse parameters becomes increasingly in demand in recent times in the rapidly growing field of coherent control to manipulate light-matter interactions.

**Acknowledgements**
This work was financially supported by Government of Russian Federation, Grant 074-U01 and Russian Foundation for Basic Research, Grant No. 16-02-00762. I.V.B. is thankful for the support of German Research Foundation (BA 4156/4-1) and Volkswagen Foundation (Nieders. Vorab. ZN3061).